# Injection and extraction magnets: septa

*M.J. Barnes*, J. Borburgh, B. Goddard, M. Hourican
CERN, Geneva, Switzerland

**Abstract**
An accelerator has limited dynamic range: a chain of accelerators is required to reach high energy. A combination of septa and kicker magnets is frequently used to inject and extract beam from each stage. The kicker magnets typically produce rectangular field pulses with fast rise- and/or fall-times, however the field strength is relatively low. To compensate for their relatively low field strength, the kicker magnets are generally combined with electromagnetic septa. The septa provide relatively strong field strength but are either DC or slow pulsed. This paper discusses injection and extraction systems with particular emphasis on the hardware required for the septa.

## 1 Introduction

An accelerator has limited dynamic range: a chain of accelerators is required to reach high energy. Thus beam transfer into (injection) and out of (extraction) an accelerator is required. The design of the injection and extraction systems aims to achieve the following:

- minimize beam loss,
- place the newly injected or extracted particles onto the correct trajectory, with the correct phase space parameters.

A combination of septa and kicker magnets is frequently used for injection and extraction. Septa can be electrostatic or magnetic: they provide slower field rise- and fall-times, but stronger field than kicker magnets. Some septa are designed to be operated with DC. Kicker magnets provide fast field rise- and fall-times, but relatively weak fields.

This paper of the CERN Accelerator School discusses the processes of injection and extraction as well as the hardware associated with the septa. The hardware associated with the kicker magnets is covered in the paper *Injection and extraction magnets: kicker magnets*.

In general, a septum (plural: septa) is a partition that separates two cavities or spaces. In a particle accelerator a septum is a device which separates two field regions. Important features of septa are an ideally homogeneous (electric or magnetic) field in one region, for deflecting beam, and a low fringe field (ideally zero magnetic and electric field) next to the septum so as not to affect the circulating beam. Hence a septum provides a space separation of circulating and injected/extracted beam. By contrast a kicker magnet provides time selection (separation) of beam to be injected/extracted.

## 2 Injection

Injection is the process of transferring a particle beam into a circular accelerator or accumulator ring, at the appropriate time, while minimizing beam loss and placing the injected particles onto the correct trajectory, with the correct phase space parameters. Injection methods include

- single-turn (fast) hadron injection,

- multi-turn hadron injection,
- charge-exchange H⁻ injection,
- lepton injection.

## 2.1 Single-turn (fast) injection

Figure 1 shows an example of fast single-turn injection in one plane. The injected beam passes through the homogeneous field region (gap) of the septum: circulating beam is in the field-free region of the septum (i.e., space separation of injected and circulating beam). The septum deflects the injected beam onto the closed orbit at the centre of the kicker magnet; the kicker magnet compensates the remaining angle. The septum and kicker are either side of a quadrupole (defocusing in the injection plane) which provides some of the required deflection and minimizes the required strength of the kicker magnet.

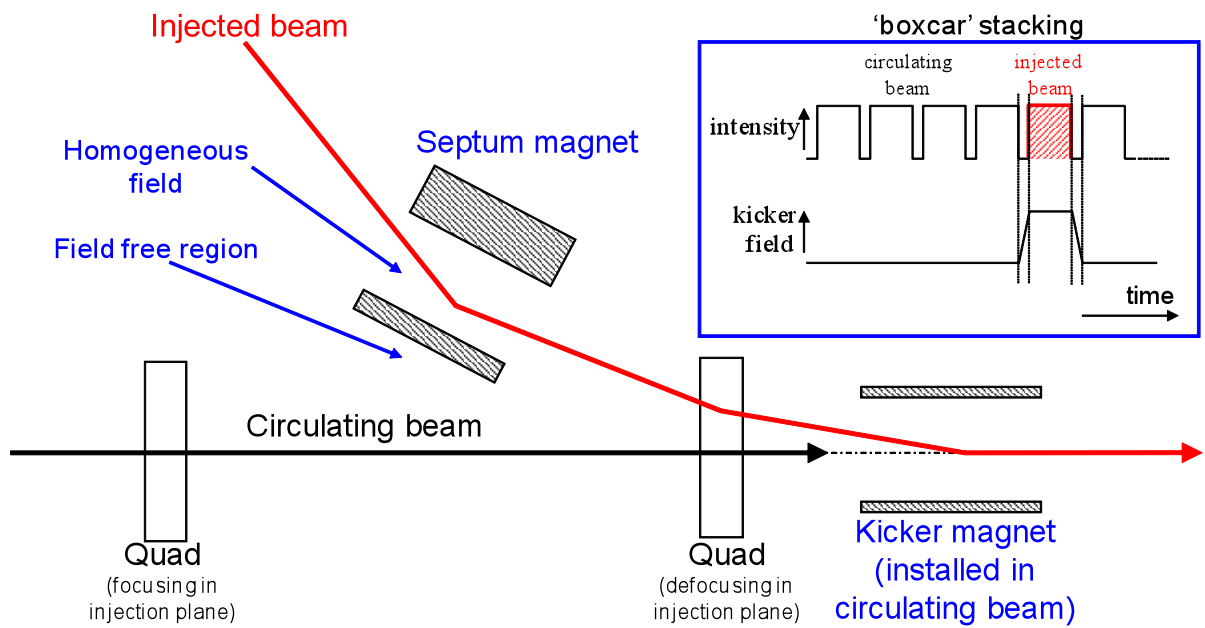

**Fig. 1:** Fast single-turn injection in one plane

The injected beam passes through the high field region of the septum once only and thus the degree of field homogeneity is not as critical as for a dipole magnet installed in an accelerator ring. However, the circulating beam passes through the 'field-free' region many times, so that achieving a very low level of stray field in the 'field-free region' is important.

The kicker magnet is installed in the accelerator and hence the circulating beam is in the aperture of the kicker. Thus the kicker field must rise from zero to full field in the time interval between the circulating beam and the start of the injected beam (Fig. 1, top right) and fall from full field to zero field in the time interval between the end of the injected beam and the subsequent circulating beam (Fig. 1, top right). The kicker magnet is discussed in more detail in the proceedings of this CERN Accelerator School, in the paper *Injection and extraction magnets: kicker magnets.*

Figure 2 shows an example of fast single-turn injection in two planes: a Lambertson septum (see Section 5.2.4) is used for a two-plane injection scheme. The injected beam passes through the homogeneous field region of the septum: circulating beam is in the field-free region of the septum. In the example shown in Fig. 2 the septum deflects the beam horizontally to the centre of the kicker magnet; the kicker magnet deflects the beam vertically onto the closed orbit. The septum and kicker are either side of an F-quadrupole (horizontally focusing and vertically defocusing) to minimize the required strength of the kicker magnet.

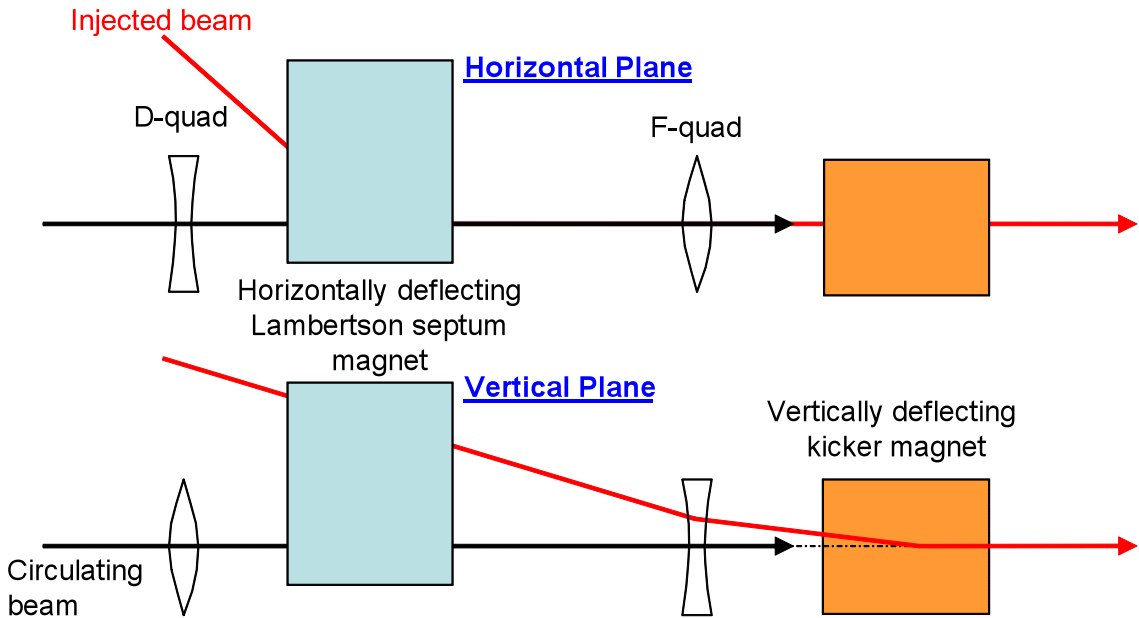

Fig. 2: Fast single-turn injection in two planes

## 2.2 Multi-turn hadron injection

A simple multi-turn injection employs a programmed orbit bumper and a septum. The orbit bumper usually bumps the beam in the horizontal plane because the horizontal acceptance is larger than the vertical acceptance in a conventional accelerator ring [1]. For hadrons the beam density at injection can be limited either by space charge effects or by the injector capacity. If the charge density cannot be increased, the horizontal phase space can sometimes be filled to increase the overall injected intensity: however, this requires the condition that the acceptance of the receiving machine be larger than the delivered beam emittance [2].

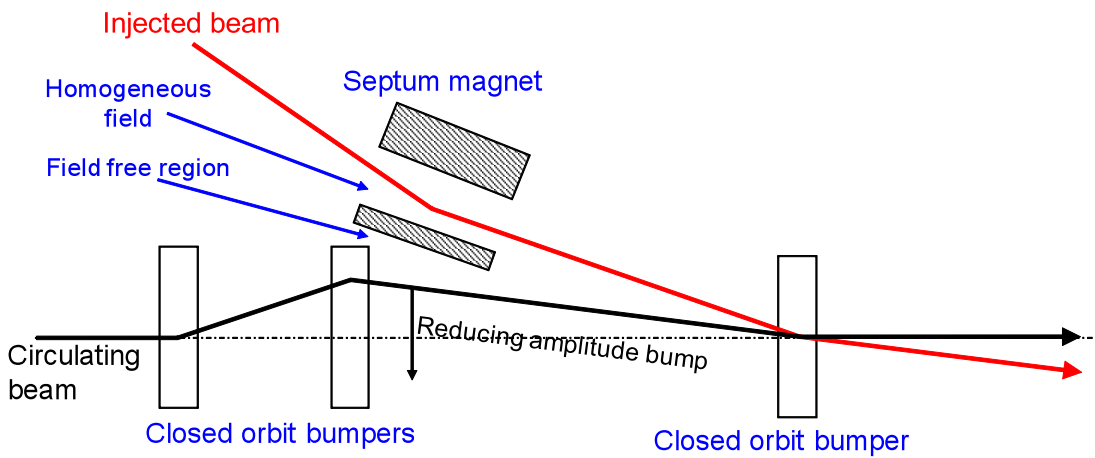

Fig. 3: Multi-turn hadron injection

Figure 3 shows an example of multi-turn hadron injection: no kicker magnet is required. The orbit bump is reduced with time so that the early beam occupies the central region of the horizontal acceptance and the later beam the periphery of the acceptance: this technique is known as phase space painting. At the end of the injection the beam bump is reduced to zero.

## 2.3 Charge-exchange H⁻ injection

Multi-turn injection is essential to accumulate high intensity. Disadvantages inherent in using an injection septum include [2]

- septum thickness of several millimetres reduces aperture,
- beam losses resulting from circulating beam hitting the septum render it radioactive,
- number of injected turns limited to 10–20.

Charge-exchange injection provides an elegant alternative method of injection. A uniform transverse phase space density is painted by modifying a closed orbit bump and steering injected beam [1, 2] (Fig. 4 and Fig. 5). The conversion from H⁻ ion to p+ means that the protons can be accumulated into already-occupied phase space, which allows very high densities to be achieved with relatively low losses.

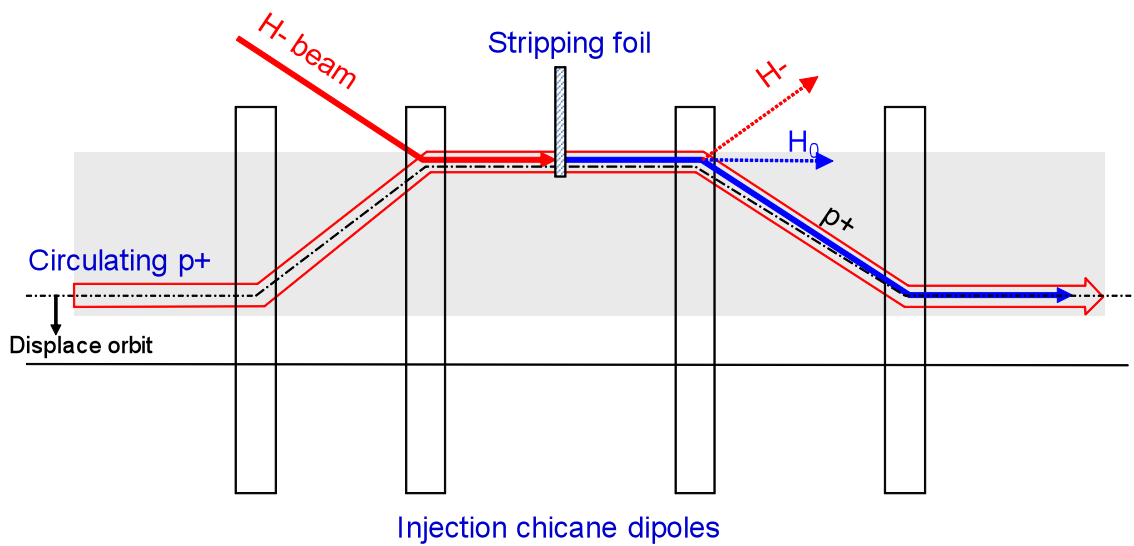

Fig. 4: Charge exchange – start of injection process

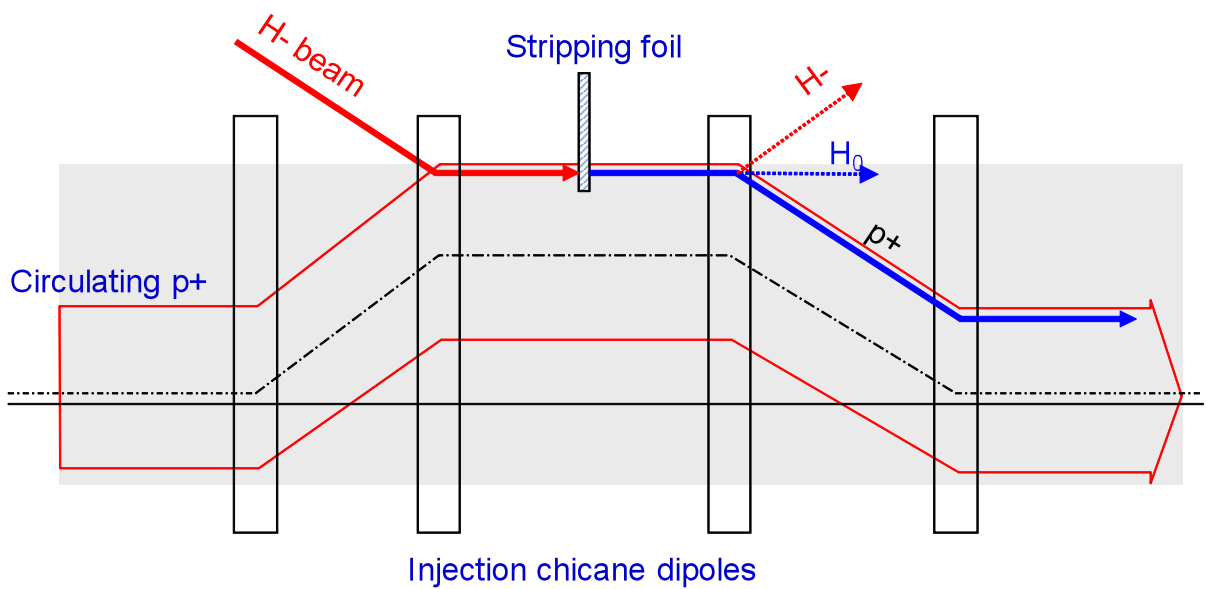

Fig. 5: Charge exchange – end of injection process

## 2.4 Lepton injection

Single-turn injection can be used as for hadrons; however, the lepton motion is strongly damped, which is not the case for proton or ion injection [2]. For lepton injection beam is injected at an angle with respect to the closed orbit (Fig. 6) and the injected beam performs damped betatron oscillations about the closed orbit [2].

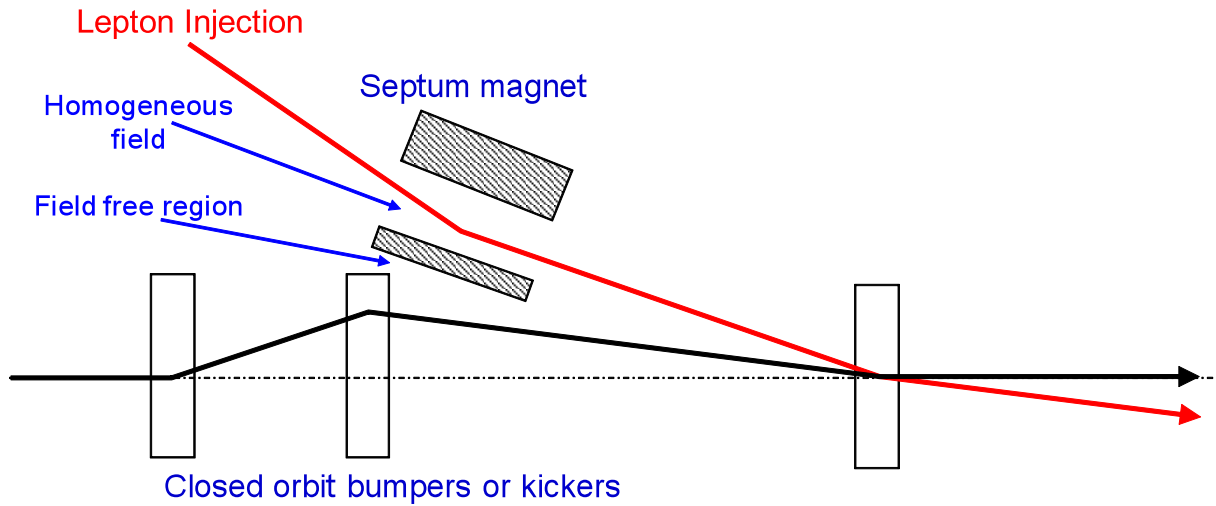

**Fig. 6:** Damped betatron lepton injection

## 3 Extraction

Extraction is the process of removing a particle beam from an accelerator to a transfer line or a beam dump, at the appropriate time, while minimizing beam loss and placing the extracted particles onto the correct trajectory, with the correct phase space parameters. Extraction methods include

– single-turn (fast) extraction,
– non-resonant multi-turn extraction,
– resonant multi-turn (slow) extraction.

Extraction usually occurs at higher energy than injection, hence stronger elements (larger $\int B.dl$) are required. At high energies many kicker and septum modules may be needed. To reduce the required strength of the kicker magnet, the beam can be moved near to a septum by a closed orbit bump.

### 3.1 Single-turn (fast) extraction

Figure 7 shows an example of fast single-turn extraction in one plane. The kicker magnet deflects the entire beam into the septum gap in a single turn (kicker magnet provides time selection [separation] of beam to be extracted). The septum deflects the entire kicked beam into the transfer line (septum provides space separation of circulating and extracted beam). The extracted beam passes through the homogeneous field region of the septum: the circulating beam, prior to extraction, is in the field-free region of the septum. A closed orbit bump can be applied to bring the circulating beam near to the septum to minimize the required strength of the kicker magnet.

As for injection, the extracted beam passes through the homogeneous field region of the septum once only and thus the degree of homogeneity is not as critical as for a dipole magnet installed in an accelerator ring. However, the circulating beam passes through the 'field-free' region many times, so that achieving a very low level of stray field in the 'field-free region' is important.

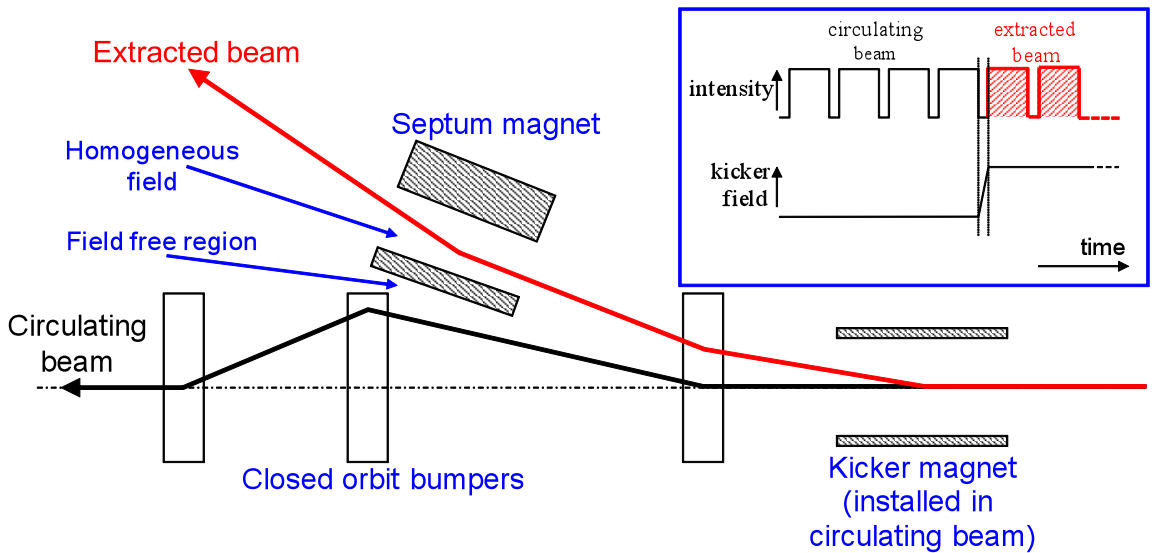

**Fig. 7:** Fast single-turn extraction in one plane

The kicker magnet is installed in the accelerator and hence the circulating beam is in the aperture of the kicker. Thus the kicker field must rise from zero to full field in a beam-free time interval deliberately created in the circulating beam (Fig. 7, top right). The entire beam is generally extracted and hence fast fall-time is typically not required: however, sometimes, bunch-by-bunch transfers are made and then the field of the kicker magnets must have fast rise- and fall-times [3]. The kicker magnet is discussed in more detail in the proceedings of this CAS, in the paper *Injection and extraction magnets: kicker magnets*.

### 3.2 Non-resonant multi-turn extraction

Some filling schemes require a beam to be injected in several turns to a larger machine. Non-resonant multi-turn extraction (over a few turns) was used, for example, for filling the SPS, at CERN, with beam extracted from the PS, with high-intensity proton beams (>2.5 $10^{13}$ protons) [2]. The process is shown schematically in Fig. 8: a fast bumper deflects the beam onto the septum and part of the beam is 'shaved' off at each turn. The beam is extracted in a few turns, with the machine tune rotating the beam. This is intrinsically a high-loss process — and hence a thin septum is essential.

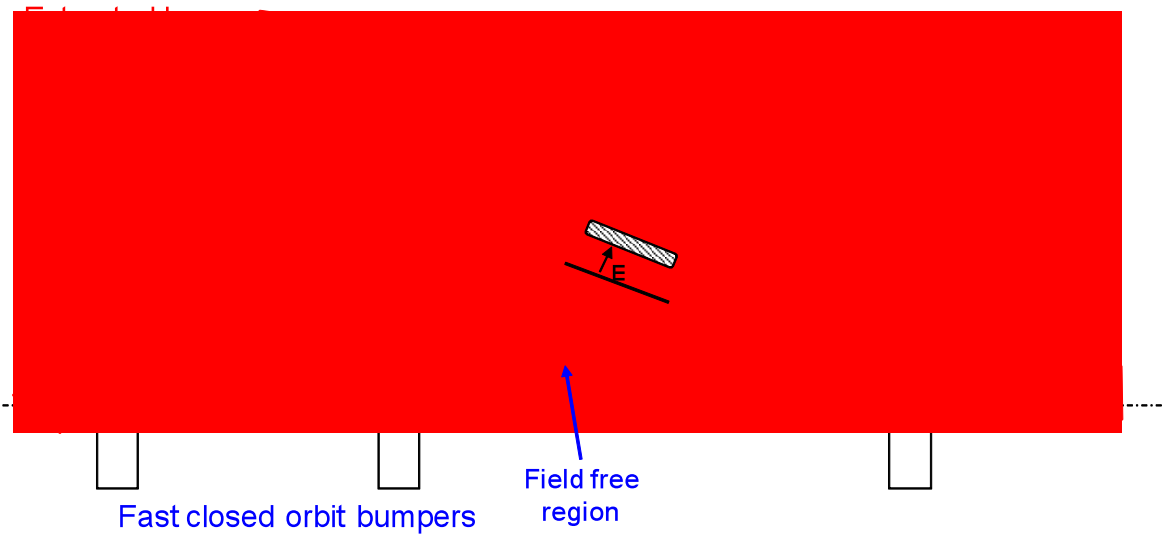

**Fig. 8:** Non-resonant multi-turn extraction

### 3.3 Resonant multi-turn (slow) extraction

Resonant multi-turn extraction is generally used for delivering beam to experiments; the extraction process can be spread over a time-interval from milliseconds to hours. Non-linear fields (slow bumpers) excite betatron resonances which drive the beam slowly across the septum (Fig. 9): this is often a third-order resonance [4].

Sextupole fields distort the circular normalized-phase-space particle trajectories: a stable area is defined, delimited by unstable fixed points [2]. Sextupole families are arranged to produce suitable phase space orientation of the stable triangle at a thin electrostatic septum. The stable area can be reduced by increasing the sextupole strength or, more easily, by approaching machine tune $Q_h$ to a resonant 1/3 integer tune [2].

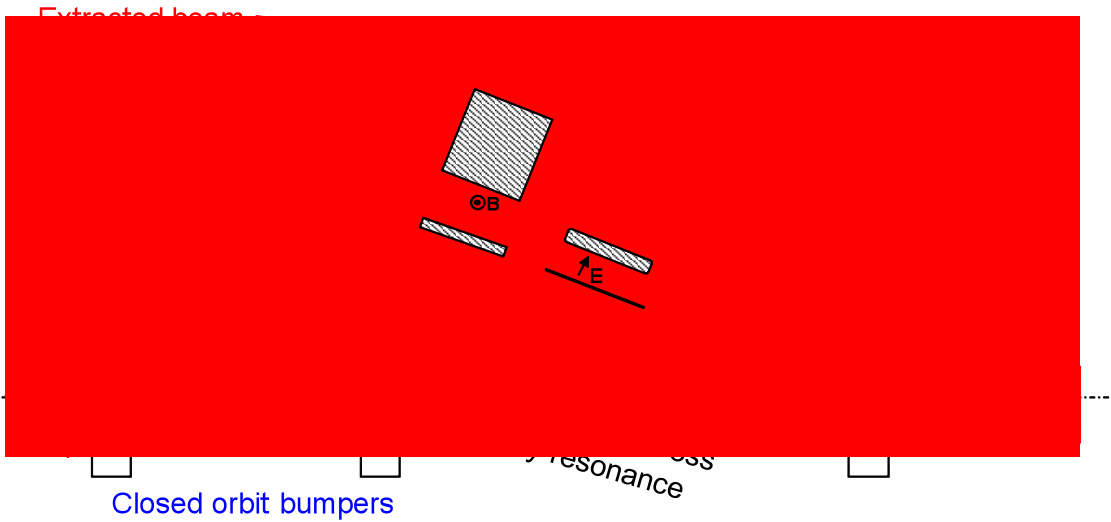

**Fig. 9:** Resonant multi-turn (slow) extraction

## 4 Deflection of beam

The force used to direct a charged particle beam is known as the Lorentz force. The Lorentz force is given by Eq. (1):

$$F = q\left[E + (v \times B)\right] , \tag{1}$$

where
- $F$ is the force (N),
- $E$ is the electric field (V/m),
- $B$ is the magnetic field (T),
- $q$ is the electric charge of the particle (C),
- $v$ is the instantaneous velocity of the particle (m/s),
- $\times$ is the vector cross product.

The deflection of a charged particle beam in a magnetic field is given by Eq. (2) [5]:

$$\theta_{B,x} = \left[\frac{0.2998}{p}\right] \cdot \int_{z_0}^{z_1} |B_y| \, dz = \left[\frac{0.2998 \cdot l_{eff}}{p}\right] \cdot |B_y| , \tag{2}$$

where

$B_y$   is the magnetic flux density in the *y*-direction (T),
*p*     is the beam momentum (GeV/*c*),
$l_{eff}$ is the effective length of the magnet [usually different from the mechanical length, due to fringe fields at the ends of the magnet] (m), and
$\theta_{B,x}$ is the deflection angle, in the *x*-direction, due to magnetic field $B_y$ (radians).

The deflection of a charged particle beam in an electric field is given by Eq. (3) [5]:

$$\theta_{E,x} = \tan^{-1}\left[\frac{1}{(p \cdot 10^9) \cdot \beta} \cdot \int_{z_0}^{z_1}|E_x|dz\right] = \tan^{-1}\left[\frac{|E_x| \cdot l_{eff}}{(p \cdot 10^9) \cdot \beta}\right] = \tan^{-1}\left[\frac{|V| \cdot l_{eff}}{d \cdot (p \cdot 10^9) \cdot \beta}\right], \quad (3)$$

where

*V*     is the potential difference between plates (V),
*d*     is the separation of the plates (m),
$E_x$   is the electric field in the *x*-direction (V/m),
*β*     is a unit-less quantity that specifies the fraction of the speed of light at which the particles travel (*v/c*), and
$\theta_{E,x}$ is the deflection angle, in the *x*-direction, due to electric field $E_x$ (radians).

## 5   Septa

Two main types of septa exist, namely electrostatic septa and magnetic septa:

– an electrostatic septum is a DC electrostatic device with very thin (typically ≤ 100 µm) separation between the zero field and high field regions,

– a magnetic septum is either a pulsed or DC dipole magnet with a thin (typically 2 mm to 20 mm) separation between the zero-field and high-field regions.

One of the significant challenges associated with the design of a septum is to achieve a low leakage field next to the septum, to avoid affecting the circulating beam, and the required field homogeneity in the gap for deflecting beam.

In order to minimize losses during the extraction process, the goal is to construct a magnetic or electrostatic septum with as thin a septum as possible. The thinnest septa are of the electrostatic type: beam is deflected by accelerating the beam perpendicular to the initial beam direction using an electric field. Using a septum which is as thin as possible increases extraction efficiency, reduces activation of the equipment and minimizes the strengths required for other extraction elements such as kicker magnets and/or preceding septa.

### 5.1   Electrostatic septum

To achieve a slow-extraction efficiency of greater than 98%, the effective thickness of the first septum unit must be ≤ 100 µm [3]. This may be realised by a very carefully aligned electrostatic septum: the septum can be a foil or an array of wires.

Figure 10 shows an electrostatic septum with a foil septum: the thin septum foil results in small interaction with beam. The orbiting beam passes through the hollow foil support, which is a field-free region. The extracted beam passes just on the other side of the septum foil (high, homogeneous, field region). Electrostatic septa use vacuum as an insulator, between septum foil/wires and electrode, and are therefore in a vacuum tank. To allow precise matching of the septum position with the circulation beam trajectory, there is often a displacement system which allows parallel and angular movement with respect to the circulating beam. The foil (or wire array — see below) is tensioned: this helps to prevent any sagging under the heat load resulting from collisions of intercepted beam particles [3].

The power supply for an electrostatic septum is typically a DC Cockroft–Walton type high-voltage generator [4].

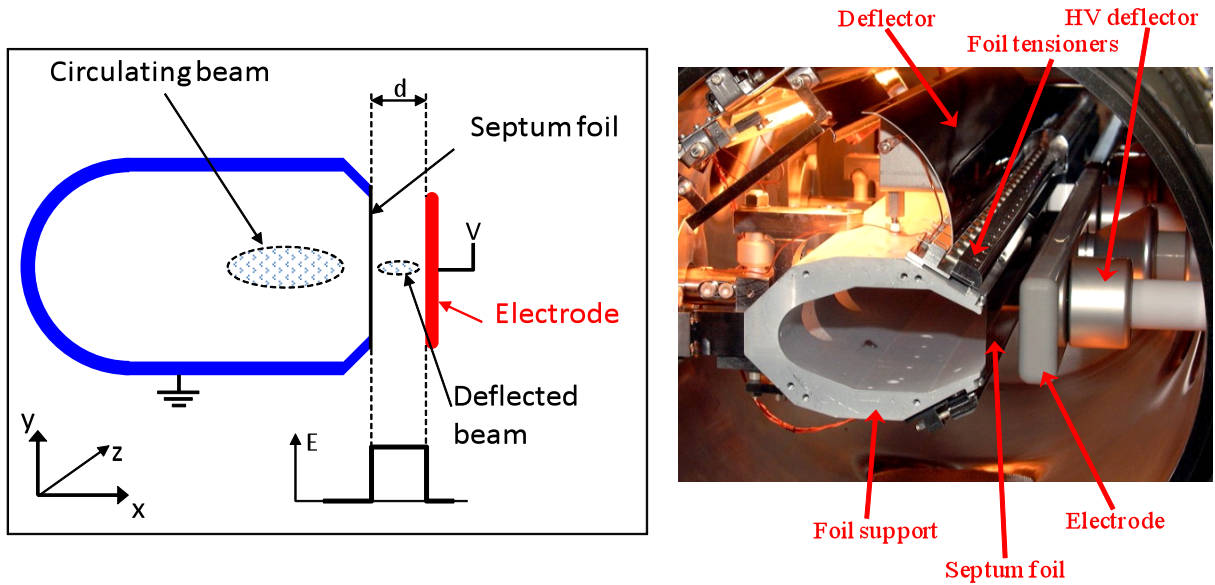

**Fig. 10:** Electrostatic septum with a foil septum

Typical technical data for an electrostatic septum include
- electrode length in the range 500 mm to 3000 mm;
- gap width variable in the range 10 mm to 35 mm;
- septum thickness of $\leq 100$ μm;
- vacuum in the range $10^{-9}$ mbar to $10^{-12}$ mbar;
- voltage up to 300 kV;
- electric field strength up to 10 MV/m;
- septum foil of molybdenum (or tungsten-rhenium wires);
- electrode made of anodized aluminium, stainless steel or, for extremely low vacuum applications, titanium;
- some electrostatic septa are bake-able up to 300°C to achieve vacuum in the $10^{-12}$ mbar range (not applicable to an aluminium electrode).

A bake-out system is required for Ultra-High Vacuum (UHV) applications. In Europe UHV is generally defined to be between $10^{-6}$ mbar and $10^{-12}$ mbar.

Conditioning and preparation of surfaces exposed to High Voltage (HV) is a significant challenge.

An adjustable foil position and gap width are useful for
- precisely adjusting position (~100 μm) and angle of the very thin septum foil to the beam position;

- selective conditioning of septum (e.g., increase gap width and increase voltage to condition feedthroughs, etc);
- permitting compensation for 'as built errors' in other equipment in the injection or extraction region.

Variants of the electrostatic septum include

- Diagonal foil: with the remote displacement at 30º from the horizontal plane. This design allows for a longitudinal painting injection scheme.
- An array of wires (Fig. 11), instead of a foil, with a diameter of ~60 μm per wire. A wire array septum allows some of the field to penetrate into the circulating beam region: the degree of penetration depends on the wire diameter and spacing. Residual gas can be ionized by the circulating beam, and the ions created can cross through the wire array into the high-field area and provoke HV breakdowns. To deal with these effects clearing electrodes, also known as ion traps, are installed (Fig. 11). These electrodes provide a vertical electric field in the circulating beam area: this electric field results in the ions being captured. At CERN, one ion trap is typically at a voltage of −3 kV and the other ion trap is typically at −7 kV: the net negative potential helps to compensate for leakage field, through the wire array, from the (negative) electrode.

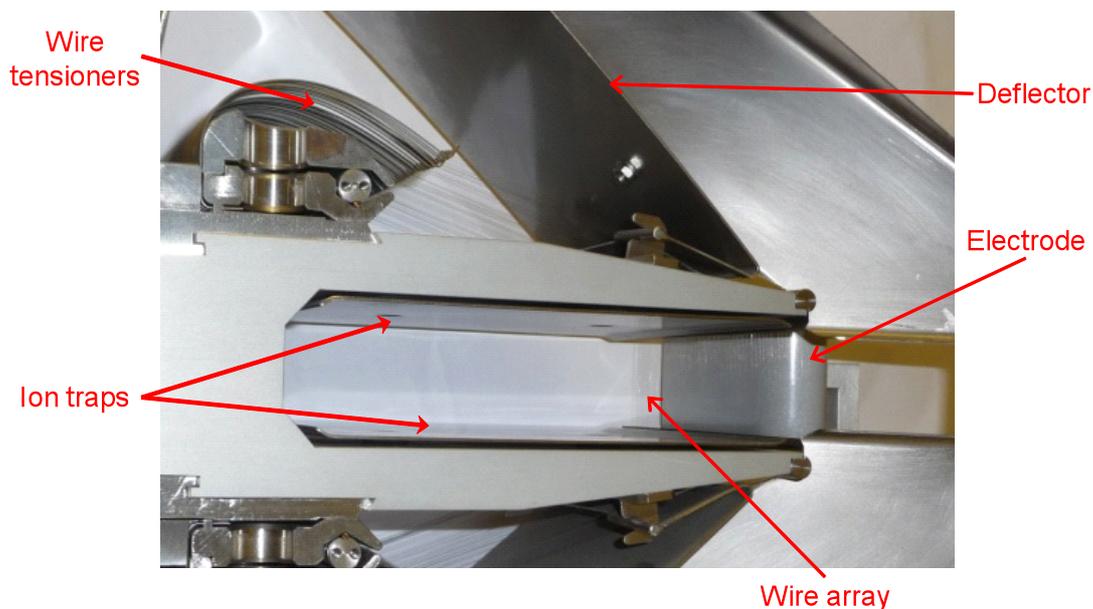

**Fig. 11:** Electrostatic septum with a wire array septum

## 5.2 Magnetic septum

The main difference between a dipole magnet and a magnetic septum is that the magnetic septum has a field-free region and a homogeneous dipole field region, separated by a relatively thin septum, whereas a dipole magnet has only a homogeneous field region. As a consequence of the relatively thin septum there is often a high current density in the septum conductor. There are several varieties of magnetic septa:

- direct-drive DC septum magnet,
- direct-drive pulsed septum magnet,
- eddy-current septum,
- Lambertson septum.

*5.2.1 Direct-drive DC septum magnet*

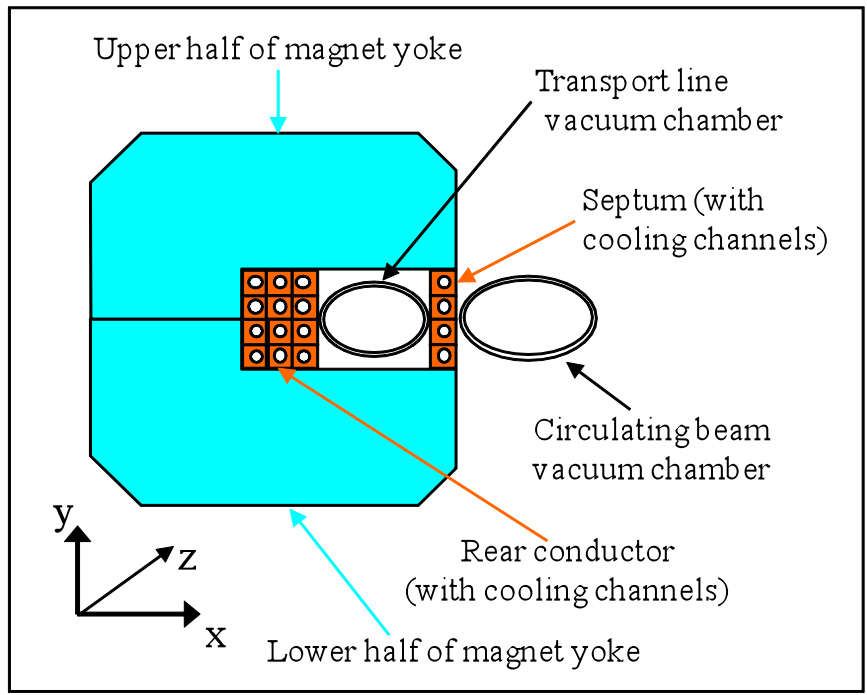

**Fig. 12:** Direct-drive DC septum magnet

Figure 12 shows a direct-drive DC septum magnet. The septum conductor is typically 6 mm to 20 mm thick: the current density in the septum conductor can be as high as 85 A/mm$^2$. The beam to be deflected passes through the gap of the septum magnet (high, homogeneous, field region): the circulating beam is on the other side of the septum conductor. A magnetic screen may be used to further reduce the leakage field into the circulating beam region.

A DC septum magnet is often used outside vacuum: in this case the coil and the magnet yoke can be split in two, an upper and a lower part, to allow the magnet to be 'clamped' over the vacuum chamber. The magnet is usually constructed with a multi-turn (series) coil, so as to reduce the current needed. However, the current required is between 0.5 kA and 4 kA and the DC septum magnet has a power consumption of up to 100 kW! Thus cooling of a DC septum is a significant issue.

Typical technical data for a direct-drive DC septum magnet are

– magnetic length per magnet yoke in the range 400 mm to 1200 mm;
– gap height of 25 mm to 60 mm;
– septum thickness of 6 mm to 20 mm;
– outside vacuum;
– laminated steel yoke;
– multi-turn coil, with water cooling circuits (flow rate: 12 l/min. to 60 l/min.);
– current in the range 0.5 kA to 4 kA;
– power supplied by controllable rectifier;
– power consumption: up to 100 kW!

The effects of insulation between turns of the conductor are discussed in Section 5.3.1.

### 5.2.2 *Direct-drive pulsed septum magnet*

Figure 13 shows a direct-drive pulsed septum magnet. This type of magnet is often used under vacuum to minimize the distance between circulating and deflected beam. The coil is generally constructed as a single-turn, so as to minimize magnet self-inductance. To allow precise matching of the septum position with the circulation beam trajectory, the magnet is also often fitted with a remote displacement system.

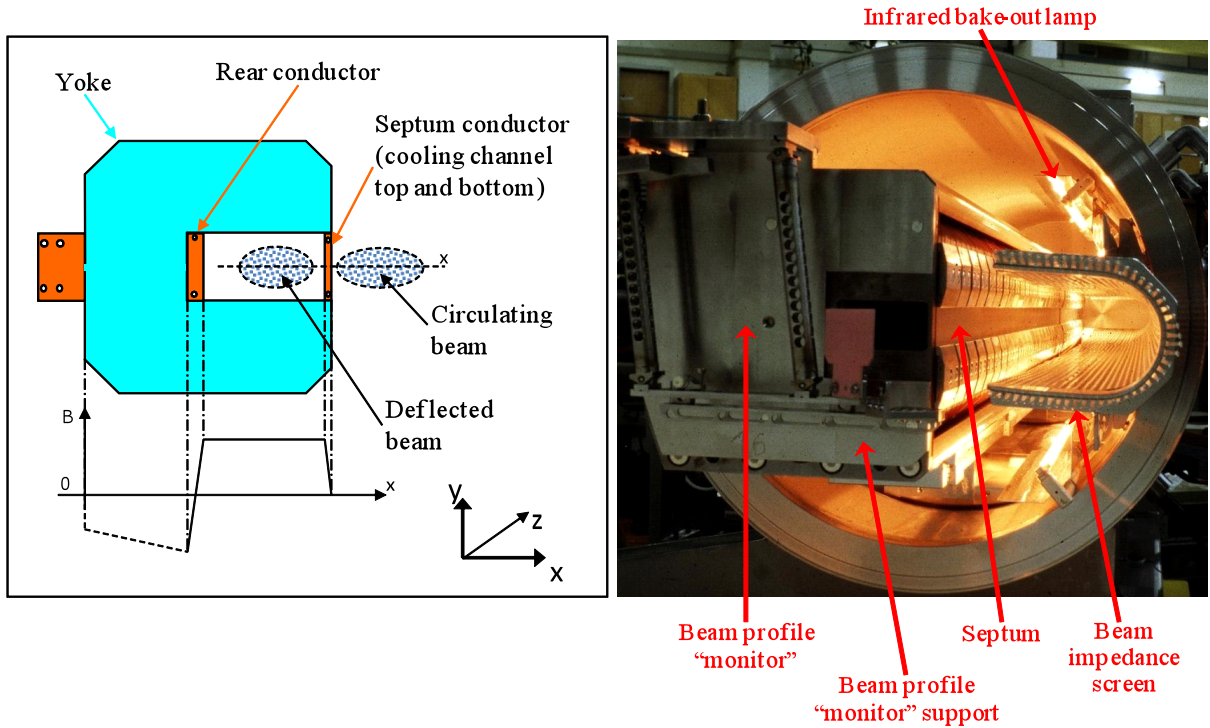

**Fig. 13:** Direct-drive pulsed septum magnet

Typical technical data for a direct-drive pulsed septum magnet are

- magnetic length per magnet yoke in the range 300 mm to 1200 mm;
- gap height of 18 mm to 60 mm;
- septum thickness of 3 mm to 20 mm;
- vacuum of ~$10^{-9}$ mbar;
- steel yoke constructed from 0.35 mm to 1.5 mm thick laminations;
- single-turn coil, with water cooling circuits (flow rate: 1 l/min. to 80 l/min.);
- bake-able up to 200°C;
- current in the range 7 kA to 40 kA, half-sine with a half-period duration of ~3 ms;
- power supplied by capacitor discharge. The flat top of the current is improved with a third harmonic circuit and active filters — (rectifier circuit used for up to 6 s 'pulse');
- a transformer is used between power supply and magnet.

High-intensity accelerators are very sensitive to longitudinal and transverse beam coupling impedance. The 'beam impedance screen', shown in Fig. 13, provides a continuous path for the image current of the circulating beam. Beam coupling impedance is discussed further in the proceedings of this CAS, in Section 4.5 of the paper *Injection and extraction magnets: kicker magnets*.

The pulsed septum magnet is powered with a half sine wave current with a half-period duration of typically 3 ms: the peak current density in the septum conductor is up to 300 A/mm$^2$. Figure 14 shows a simplified schematic for a power supply [6, 7] for powering a pulsed septum magnet; a third harmonic circuit is used to obtain a better flat top current than is given by a basic sinusoidal discharge current:

- The capacitors are accurately charged to the required voltage.
- The 'third harmonic circuit' generates a current which is superimposed upon (adds to) the discharge current from the 'fundamental circuit'.
- A transformer is used to allow the use of standard 2 kV capacitors on the primary and to give the required high current on the secondary. The transformer turns-ratio N1:N2 (Fig. 14) is typically in the range 4:1 to 50:1 [7].
- An active filter circuit (not shown) can be used to obtain a stability of flat top current of 10$^{-4}$ over a time of 500 µs [7].

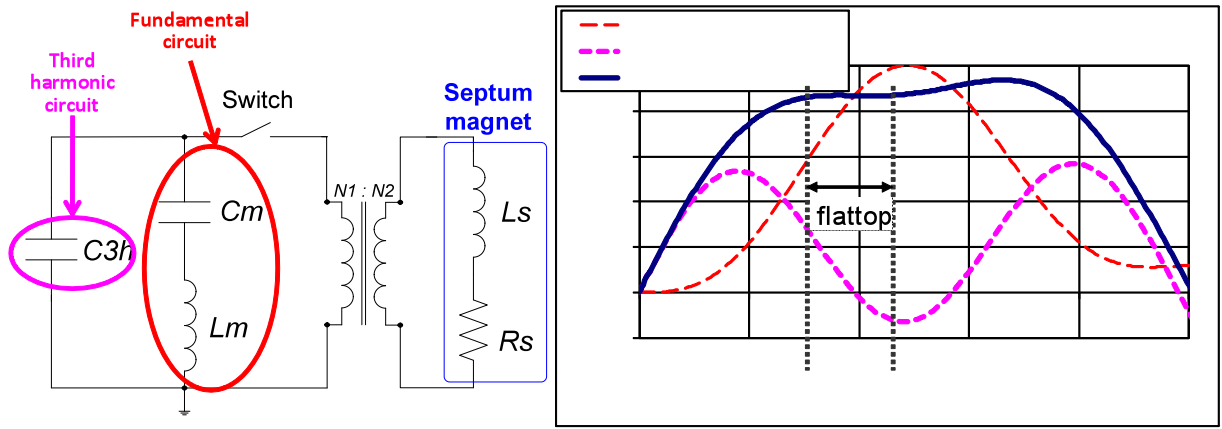

**Fig. 14:** Third harmonic circuit and example waveforms for a pulsed magnetic septum

### 5.2.3 *Eddy-current septum*

An eddy-current septum [8] (Fig. 15) is powered with a half or full sine wave current with a period of typically 50 µs. The coil is generally constructed as a single-turn, so as to minimize magnet self-inductance. The coil is situated around the back leg of the C-shaped yoke (Fig. 15), and therefore coil dimensions are generally not critical. When the magnet is pulsed, the magnetic field induces eddy-currents in the septum, counteracting the fringe field created. The septum conductor can be made thinner than for the direct drive septum, but cooling circuits may be needed at the edges to cool the septum.

The field in the septum gap as function of time follows the coil current. The electrical resistance of the septum is kept low: once the septum current is flowing, it takes quite some time to decay away. The Left Hand Side (LHS) of Fig. 15 shows an eddy-current septum without a return box and magnetic screen: a typical maximum leakage field would be 10% of the gap field. To reduce further the fringe field of the eddy-current septum a copper box (return box) can be placed around the septum magnet: this is shown in the Right Hand Side (RHS) of Fig. 15. In addition a magnetic screen can be added next to the septum conductor. These modifications permit the fringe field, seen by the circulating beam, to be reduced to below 0.01% of the gap field at all times and places [9].

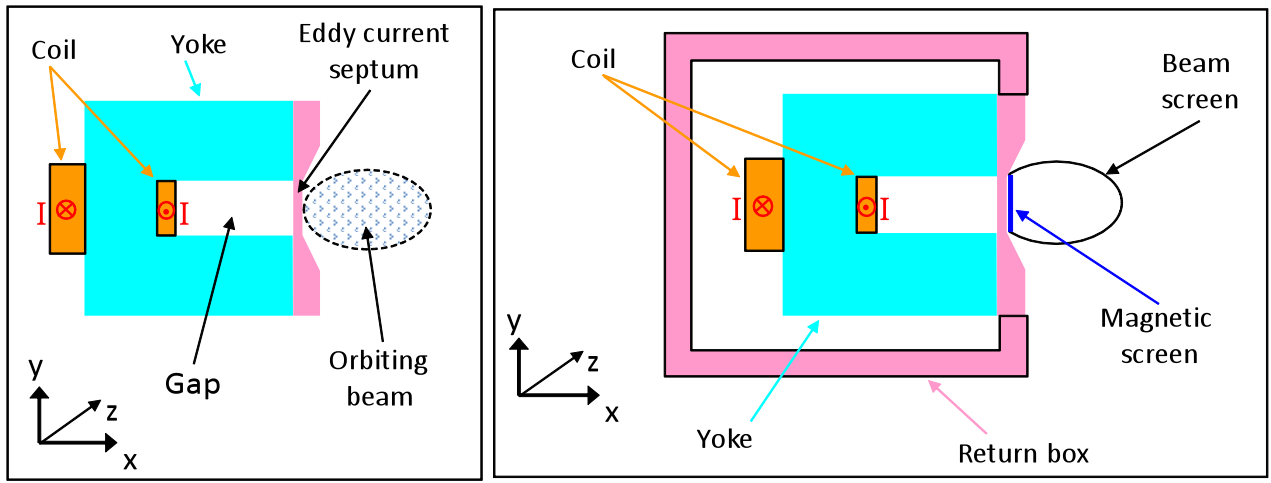

**Fig. 15:** Eddy-current septum: without return box or magnetic screen (LHS), with return box and magnetic screen (RHS)

Typical technical data for an eddy-current septum are

- magnetic length per magnet yoke in the range 400 mm to 800 mm;
- gap height of 10 mm to 30 mm;
- septum thickness of 1 mm to 3 mm;
- vacuum of ~$10^{-9}$ mbar, or out of vacuum;
- steel yoke with 0.1 mm to 0.35 mm thick laminations;
- single-turn coil, with water cooling circuits (flow rate: 1 l/min. to 10 l/min.);
- current of ~10 kA peak;
- fast pulsed with 50 μs period;
- powered with a capacitor discharge: half-sine or full-sine wave.

### 5.2.4 *Lambertson septum*

A Lambertson iron-septum is generally a fairly rugged device [10]. The conductors are enclosed in the steel yoke, 'well away' from the beam. The Lambertson septum used in the LHC injection beam line is shown in Fig. 16: this septum is a DC device but pulsed Lambertson septa also exist. For the injection into the LHC, the transfer line from the SPS passes through the gap of the Lambertson septum [10]. There is a thin steel yoke between the gap of the Lambertson septum and septum hole, containing circulating LHC beam in a beam pipe (Fig. 16) — however, sufficient steel is required to avoid saturation. As shown in Fig. 16, the septum deflects beam horizontally to the right; the downstream LHC injection kicker magnets deflect the beam vertically into the septum hole, onto the central orbit of the circulating beam. To minimize field in the septum hole and beam hole (Fig. 16), containing circulating beams, an additional screen is used around each LHC beam pipe.

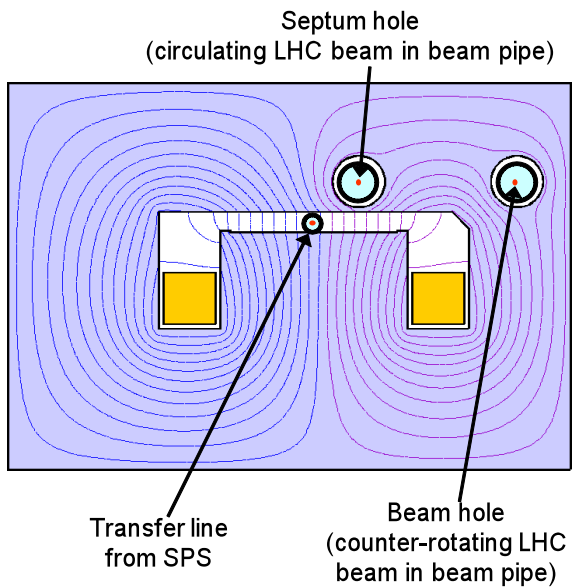 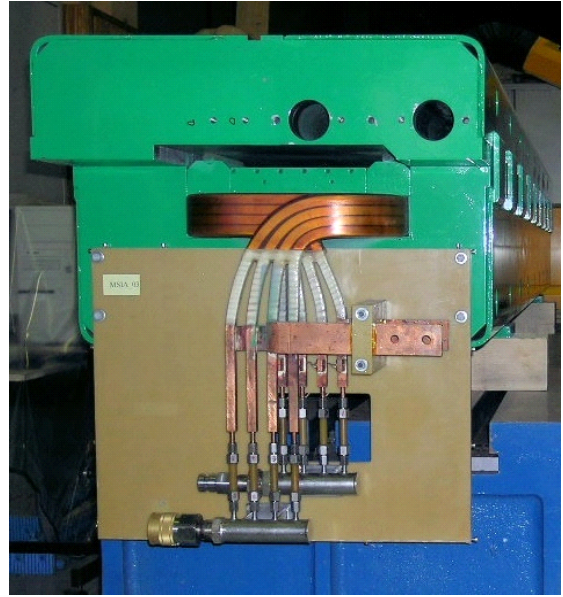

**Fig. 16:** Lambertson septum used in LHC injection beam-line

## 5.3 Practical considerations

### 5.3.1 *Insulation and cooling of septum magnet*

A single-turn coil may be inserted (with minimum clearance) into the gap at the outer end of a C-type yoke (Fig. 17). If the core permeability is high and the current sheet has a nearly uniform current density, there is little leakage flux outside the gap, except at the magnet ends, and excellent field homogeneity is achieved in the aperture.

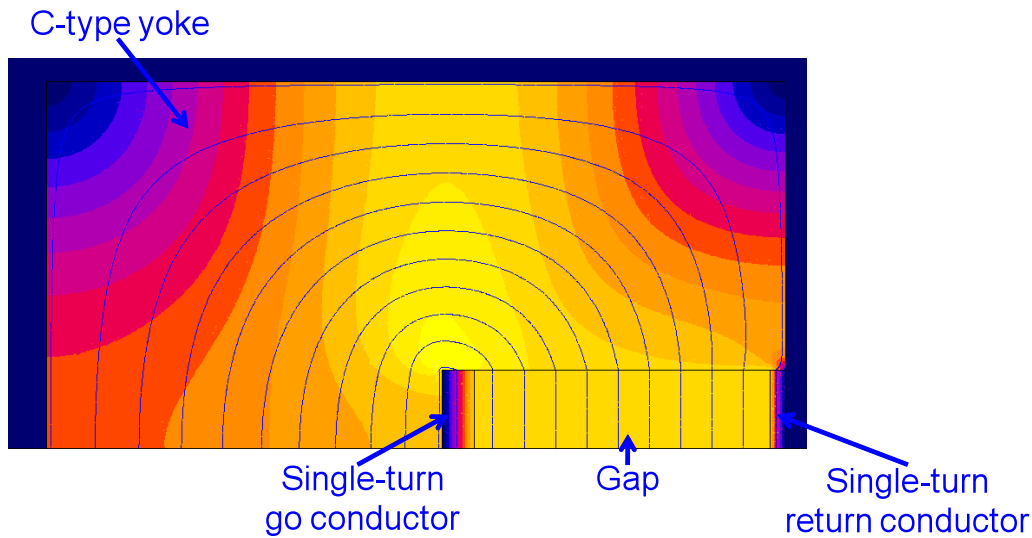

**Fig. 17:** 'Idealized' septum magnet; minimum clearance between single-turn coil and C-type yoke

However, in reality, there is a small space between the coil and yoke which is not occupied by conductor. In addition, a multi-turn septum has, by necessity, insulation between the individual windings, thus the current density is not uniform. Further the presence of cooling channels (e.g., Fig. 12) affects the uniformity of the current density. The overall effect of insulation and cooling is to increase the leakage field which, if other measures are not taken, could be up to ~2% of the gap field: in addition the field inhomogeneity in the gap can be up to ±2%.

In a thin edge-cooled septum the high temperature gradient changes the local resistivity of the septum conductor and thus the uniformity of the current density: to achieve a more uniform current density, the septum is profiled to compensate for the resistivity change resulting from the temperature gradient. In addition a significant increase in the temperature of the coil and power connections can lead to load changes as seen by the power supply: thus suitable regulation of the power supply is required.

A small septum conductor thickness results in high current density, high thermal loads and high mechanical stress on the coil. Thus cooling is a major design consideration for a septum magnet. The flow characteristics (laminar/turbulent/mixed) in the cooling tube are dependent on the Reynolds number ($R_e$):

$$R_e = \left( \frac{\rho U_m D}{\mu} \right), \tag{4}$$

where
- $\rho$ is the density of the cooling fluid (kg·m$^{-3}$),
- $U_m$ is the mean velocity of the cooling fluid (m/s),
- $D$ is the diameter of the cooling tube (m), and
- $\mu$ is the dynamic viscosity of the cooling fluid (N·s·m$^{-2}$).

Excessive cooling flow rate will lead to cavitation and erosion. Thus the cooling circuit must be designed for optimum flow conditions: turbulent flow results in best heat exchange, but laminar flow results in low erosion and cavitation. An optimum flow-condition typically corresponds to a turbulent flow with water speeds up to 10 m/s [11].

### 5.3.2 Vacuum considerations of magnetic septum

To reduce septum thickness, as seen by the beam (apparent septum thickness), complex, thin-walled, vacuum chambers can be used around which an outside vacuum magnetic septum can be clamped (Fig. 18). Complicated (difficult to manufacture) and UHV compatible (material quality) vacuum chambers are often required for injection/extraction points.

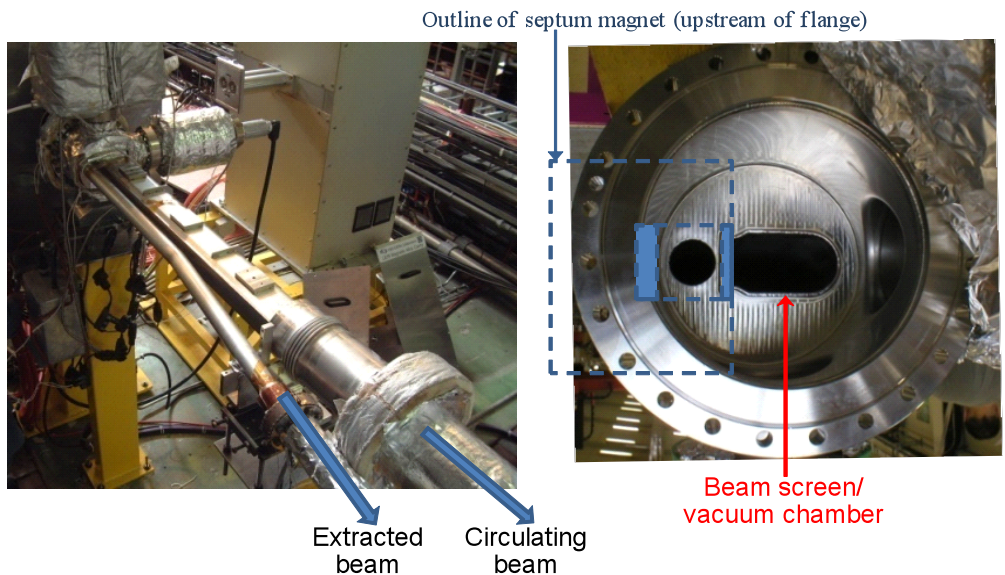

**Fig. 18:** To reduce apparent septum thickness, complex, thin-walled, vacuum chambers can be used (LHS), around which the septum magnet can be clamped (RHS)

To reduce apparent septum thickness even further, the magnetic septum can be put under vacuum. To reach UHV, pumping is required and bake-out may be necessary which requires suitable mechanical design and all the relevant heating equipment. In this case under vacuum heaters, heating jackets, reflectors, ceramic insulators, etc. are required. In some cases, where pressure is critical, a Non-Evaporable Getter (NEG) coating may have to be applied to the chambers, which requires activation systems.

### 5.3.3 Forces on a magnetic septum

The mechanical forces on the septum conductor can be significant and are normally at a maximum on the mid-axis of the septum conductor. High cycle numbers can lead to fatigue problems in thin copper cross-section. The force on the conductor ($F_b$) is given by Eq. (5):

$$F_b = (BI_s l_s / 2),\qquad(5)$$

where

- $B$ is the flux-density in the gap (T),
- $I_s$ is the current in the septum conductor (A), and
- $l_s$ is the length of the septum conductor (m).

The force on the conductor can exceed 10 kN: such force results in deflection of the conductor, of up to 40 μm, which can result in fatigue failure for a pulsed septum. The coil fixation, for a pulsed septum, is designed to be flexible and therefore uses springs (Fig. 19). The springs can be made of beryllium copper alloy and are inserted at regular intervals along the length of the coil: the spring is in contact with the septum via a lever which is then clamped in a slot in the magnet yoke (Fig. 19).

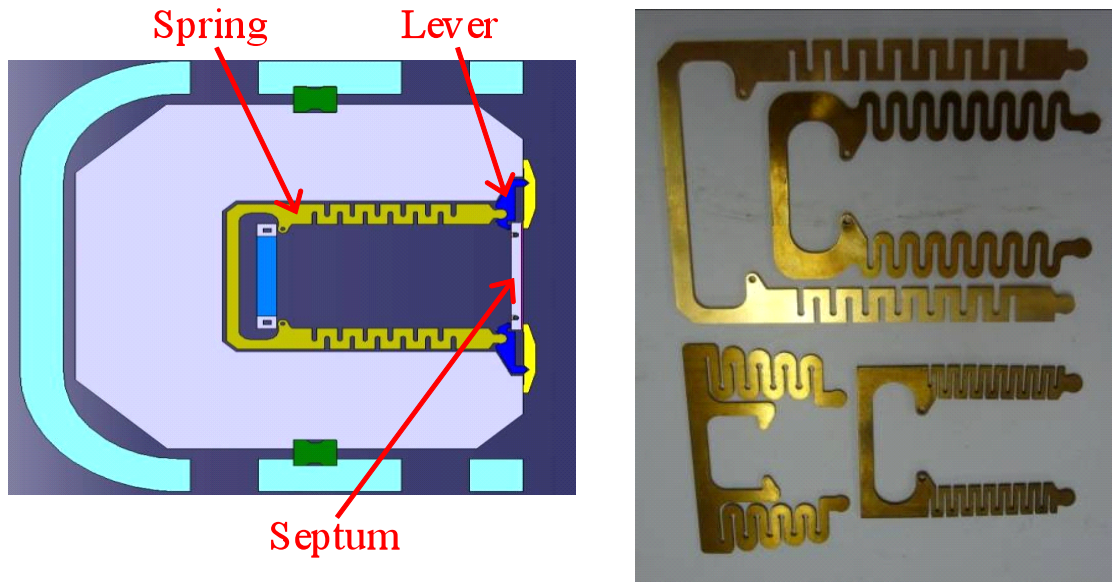

**Fig. 19:** Installation of springs (LHS) and examples of springs (RHS)

## Acknowledgements

The authors wish to acknowledge the contributions of B. Balhan, J-M. Cravero, M. Gyr, and T. Masson to this article.

# References

[1] G.H. Rees, Injection, CAS - CERN Accelerator School: 5th General accelerator physics course, Jyväskylä, Finland, 1992, CERN 94-01 v 2.

[2] B. Goddard, Transfer lines, Introduction to Accelerator Physics 2009, February 23–27, 2009, La Villa du Lac, Divonne, France, http://indico.cern.ch/conferenceDisplay.py?confId=43703

[3] G.H. Rees, Extraction, CAS - CERN Accelerator School: 5th General accelerator physics course, Jyväskylä, Finland, 1992, CERN 94-01 v 2.

[4] O. Bruning, CERN complex, Introduction to Accelerator Physics 2009, February 23–27, 2009, La Villa du Lac, Divonne, France, http://indico.cern.ch/conferenceDisplay.py?confId=43703.

[5] C. Bovet, R. Gouiran, I. Gumowski, K.H. Reich, A selection of formulae and data useful for the design of A.G. synchrotrons, CERN/MPSSI/Int. DL/70/4.

[6] J-M. Cravero and J.P. Royer, The new pulsed power converter for the septum magnet in the PS straight section 42, CERN PS/PO/ Note 97-03, 1997.

[7] J.P. Royer, High current with precision flat-top capacitor discharge power converters for pulsed septum magnets, CERN/PS 95-13 (PO), 1995.

[8] M.J. Barnes, B. Balhan, J. Borburgh, T. Fowler, B. Goddard, W.J.M. Weterings, A. Ueda, Development of an eddy current septum for Linac4, Proc. 11[th] European Particle Accelerator Conference (EPAC'08), Genoa, Italy, June 23–27, 2008, pp. 1434–1436.

[9] P. Lebasque et al., Eddy current septum magnets for booster injection and extraction and storage ring injection at synchrotron soleil, Proc. 10[th] European Particle Accelerator Conference (EPAC'06), Edinburgh, Scotland, June 26–30, 2006, pp. 3511–3513.

[10] S. Bidon et al., Steel septum magnets for the LHC beam injection and extraction, Proc. 8[th] European Particle Accelerator Conference (EPAC'02), Paris, June 3–8, 2002, pp. 2514–2516.

[11] R.L. Keizer, Calculation of DC operated septum magnets cooling problems, CERN SI/Int. MAE/71-1.

# Bibliography

J. Borburgh, M. Crescenti, M. Hourican, T. Masson, Design and construction of the LEIR extraction septum, *IEEE Trans. on Applied Superconductivity*, Vol. 16, No. 2, June 2006.

J. Borburgh, B. Balhan, T. Fowler, M. Hourican, W.J.M. Weterings, Septa and distributor developments for H- injection into the booster from Linac4, Proc. 11[th] European Particle Accelerator Conference (EPAC'08), Genoa, Italy, June 23–27, 2008, pp. 2338–2340.

Y. Yonemure et al., Beam extraction of the POP FFAG with a massless septum, Proc. 20th Particle Accelerator Conference (PAC'03), Portland, Oregon, USA, May 12–16, 2003, pp. 1679–1681.

S. Fartoukh, A semi-analytical method to generate an arbitrary 2D magnetic field and determine the associated current distribution, LHC Project Report 1012, CERN.

Y. Iwashite, A. Noda, Massless septum with hybrid magnet, Proc. 6[th] European Particle Accelerator Conference (EPAC'98), Stockholm, Sweden, June 22–26, 1998, pp. 2109–2110.

I. Sakai et al., Operation of an opposite field septum for the J-Parc main ring injection, Proc. 10[th] European Particle Accelerator Conference (EPAC'06), Edinburgh, Scotland, June 26–30, 2006, pp. 1750–1752.

M. Benedikt, P. Collier, V. Mertens, J. Poole, K. Schindl (Eds.), LHC Design Report, volume III, The LHC injector chain, http://ab-div.web.cern.ch/ab-div/Publications/LHC-DesignReport.html, CERN, Geneva, 2004, CERN-2004-003.

E. Dallago, G. Venchi, S. Rossi, M. Pullia, T. Fowler, M.B. Larsen, The power supply for the beam chopper magnets of a medical synchrotron, 37[th] IEEE Power Electronics Specialists Conference, 2006, (PESC '06), Jeju, Korea, June 18–22, 2006.